\documentclass[12pt,preprint]{aastex}

\slugcomment{Submitted to ApJ}

\shorttitle{Indications for Rotation in the  DG Tau jet}
\shortauthors{Bacciotti et al.}

\begin{document}

\title{HST/STIS Spectroscopy of the Optical  Outflow from
DG Tau: Indications for Rotation in the Initial Jet Channel$^1$}


\author{Francesca Bacciotti\altaffilmark{2,3},
Thomas P. Ray\altaffilmark{2},
Reinhard Mundt\altaffilmark{4},
\\ 
Jochen Eisl\"{o}ffel\altaffilmark{5},
Josef Solf\altaffilmark{5}}

\altaffiltext{1}{Based on observations made with the 
NASA/ESA {\em Hubble Space Telescope}, obtained at 
the Space Telescope Science Institute, 
which is operated by the Association 
of Universities for Research in Astronomy,
Inc., under NASA contract NAS5-26555.} 
\altaffiltext{2}{School of Cosmic Physics, 
Dublin Institute for Advanced Studies,
5 Merrion Square, Dublin 2, Ireland.}
\altaffiltext{3}{Osservatorio Astrofisico di Arcetri, Largo E. Fermi 5,
I-50125  Florence, Italy}
\altaffiltext{4}{Max-Planck-Institut f\"ur Astronomie, K\"onigstuhl 17,
D-69117  Heidelberg, Germany}
\altaffiltext{5}{Th\"uringer Landessternwarte Tautenburg,
Sternwarte 5, D-07778, Tautenburg, Germany}

\begin{abstract}
We have carried out a kinematical, 
high angular resolution ($\sim$ 0\farcs 1) study of the
optical blueshifted flow from DG~Tau 
within 0\farcs 5 from the source 
(i.e.\ 110 AU when de-projected along this flow).
We analysed optical emission line profiles 
extracted from a set of seven long-slit spectra taken 
with the  {\em Space Telescope Imaging Spectrograph} 
(STIS) on board the Hubble Space Telescope (HST),
obtained by maintaining the slit parallel to the 
outflow axis while at the same time moving it transversely in steps of 
0\farcs 07. For the spatially resolved flow of moderate 
velocity (peaking at -70\,km\,s$^{-1}$),
we have found systematic differences in the radial velocities of lines
from opposing slit positions i.e.\ on alternate sides of the
jet axis. The results, obtained using two
independent techniques, are corrected for the spurious 
wavelength shift due to the uneven 
illumination of the STIS slit. Other instrumental effects 
are shown to be either absent or unimportant. 
The derived relative Doppler shifts range 
from 5 to 20\,km\,s$^{-1}$. Assuming the flow is axially symmetric,
the velocity shifts are consistent with the southeastern side of the 
flow moving towards the observer faster than the corresponding 
northwestern side. If this finding is interpreted
as rotation, the flow is then rotating clockwise looking 
from the jet towards the source and the derived toroidal velocities
are in the range  6 to 15\,km\,s$^{-1}$,
depending on position.
Combining these values with recent estimates of the 
mass loss rate, one would obtain 
an angular momentum flux, for the low to moderate velocity 
regime of the flow, of 
$\dot{J}_{w,lm}\sim 3.8\,10^{-5}$\,M$_\odot$\,yr$^{-1}$ AU\,km\,s$^{-1}$.
Our findings may constitute 
the first detection of rotation in the initial channel
of a jet flow. The derived values appear to be consistent 
with the predictions of popular magneto-centrifugal jet-launching 
models, although we cannot exclude the possibility that the 
observed velocity differences are due to some transverse outflow 
asymmetry other than rotation.   
\end{abstract}



\section{Introduction} \label{intro}

Herbig-Haro (HH) jets, optically emitting collimated mass outflows associated 
with young stellar objects (see, e.g., \citet{ray98}, \citet{eis00}), 
are widely recognised as an essential ingredient of the star formation 
process. 
In particular, they are believed to contribute to 
the removal of excess angular momentum from accreted matter and to  
disperse infalling circumstellar envelopes. 
Despite their possible key role in star formation,  
the origin of the jets themselves remains elusive, although 
it is believed that their generation involves
the simultaneous action of magnetic and centrifugal forces 
in a rotating star/disk system \citep{konigl00,shu00,shib99}.
Canonical models, however, have not been tested observationally,
since the process is believed to occur on very small scales 
(less than a few AU), although according to some, the acceleration
and collimation region may extend to $\sim$\,100\,AU from the star. 

With these ideas in mind, we have observed  DG~Tau with
STIS on-board the HST. Multiple 
overlapping slit positions parallel to the outflow 
from this star were chosen 
so as to  build up a 3-D spatial intensity-velocity ``cube''  
in various forbidden emission lines (FELs) and H$\alpha$. 
In a previous paper \citep{bmresc00} we have presented 
high spatial resolution ($\sim$ 0\farcs 1) velocity channel maps 
of the first 2\arcsec\ of this flow in different 
emission lines and in several distinct radial velocity intervals. 
In a subsequent paper \citep{brmes02} we will present 
mono- and bi-dimensional maps of the excitation and dynamical 
properties of the same region of the flow, 
also in various velocity intervals. 
Briefly the main conclusions of these studies are as follows. 
The outflow appears to have an onion-like kinematic structure,
with the faster and more collimated gas 
continuously bracketed by wider and slower 
material. In addition, the flow becomes gradually denser, 
and of higher excitation, with proximity to the central axis. 
Combining these results, we have been
able to calculate a flow mass loss rate, 
$\dot{M}_w \sim 2.4~10^{-7}$\,M$_\odot$\,yr$^{-1}$, which is 
about one tenth of the estimated mass accretion rate through the disk
\citep{heg95,brmes02}.
These results are in line with what is predicted by
popular magneto-hydrodynamic (MHD) models. 

Further evidence, however, that
would help to support such models is observation 
of rotation in the initial section of the flow. 
In the magneto-centrifugal scenario, the flow maintains a record of rotation
at its base, at least initially, during propagation. 
If the flow has a favourable inclination angle with respect
to the line of sight, a trace of its rotation should be seen 
in sets of high angular resolution spectra taken 
close to the source with the slit parallel to the outflow axis.
Our DG~Tau dataset is ideally suited for such an investigation. We note that 
hints of rotation were found in the HH~212 flow at large distances 
(2\,10$^3$ to 10$^4$\,AU) from the source by \citet{dbsch00}.
Although these results are very encouraging, we nevertheless 
believe that the launching mechanism is better constrained by 
the kinematical properties of the {\em initial portion} 
of the jet channel. Here, the flow may not have suffered the effects of 
strong interactions with its environment. 

The results of our analysis 
are described in Section~\ref{rot}, after a brief summary of
our observations and data reduction (Section~\ref{obs}). 
We then discuss our findings in Section~\ref{disc} in the 
light of MHD jet models.

\section{Observations and data reduction} \label{obs}

Our HST/STIS observations and calibration 
procedures were described in detail in \citet{bmresc00}. 
Briefly, our seven STIS spectra of the DG~Tau flow, 
taken with the G750M grating, 
included the brightest forbidden lines, 
[OI]\,$\lambda\lambda$6300,6363, 
[NII]\,$\lambda\lambda$6548,6583, 
[SII]\,$\lambda\lambda$6716,6731 along with H$\alpha$.
The slit aperture was 52$\times$0.1\,arcsec$^2$, while
the spectral resolution was  0.554\,\AA\,pixel$^{\rm -1}$.
The nominal spatial sampling was 0\farcs 05\,pixel$^{-1}$,
and the Point Spread Function (PSF) of HST in the red gives 
an effective angular resolution of about 0\farcs1 (FWHM).
The slit was kept parallel to the outflow axis (P.A. 226\degr), 
and offset to the southeast and northwest of the jet axis in steps of
0\farcs 07, for a total coverage in the direction 
transverse to the jet of  about  0\farcs 5. 

With the spectra taken in the seven slit positions
(labelled S1,S2,...S7 going from the southeast to 
the north-west) we formed images of the flow 
(channel maps) in four broad velocity bins, and in four lines  
(see Figs.\,1 and 2 of \citet{bmresc00}). The velocity 
range was approximately +50\,km\,s$^{-1}$ to -450\,km\,s$^{-1}$ 
and the four coarse velocity bins, with widths of about 
125\,km\,s$^{-1}$ each, are labelled here as 
low, medium, high, and very high velocity intervals  
(i.e. LVI, MVI, HVI and VHVI) respectively. 
The heliocentric radial velocity of the system, 
v$_{\star, hel} \approx$ +16.5\,km\,s$^{\rm -1}$, 
was derived from the LiI\,$\lambda$6707 
photospheric absorption line.
To search for evidence of rotation, we concentrated on the 
LVI and MVI.  In these velocity intervals the jet is 
bright close to the source (especially in the [SII] and [OI] lines), 
and laterally extended, i.e. this emission  
derives from the external part of the collimated wind.
In contrast, the highest velocity emission is highly collimated and 
concentrated towards the central axis. To  
examine the transverse velocity profile of the
latter would require an angular resolution 
beyond that currently available with HST. 

If the flow is rotating, the gas located on opposite 
sides of the  channel, with respect to the central axis, 
should emit lines having slightly different Doppler shifts.  
We thus examined the line profiles in each position along the jet,
in all of the seven spectra.
We divided the flow into four regions (labelled I -- IV) of increasing 
distance from the star, and we longitudinally averaged the emission 
from within each region. Region I is from 
0\farcs 025 to  0\farcs 125, II
from 0\farcs 125 to  0\farcs 225, III from 0\farcs 225 to  0\farcs 325 and IV
from 0\farcs 325 to 0\farcs 475 away from the source (the latter region is 
larger, because here the emission begins to be 
diluted in a `bubble' feature, see \citet{bmresc00}).

The spectra were first corrected for  the 
spurious wavelength offsets introduced in the above regions
by the uneven illumination of the STIS slit (see 
\citet{marconi02} and  Appendix~\ref{app1}).
This correction is necessary because in our observations
the transverse intensity 
distribution peaks at the location of the central slit, 
thus the spurious wavelength shifts have  reverse 
signs in slits oppositely placed with respect to the central one, 
potentially mimicking rotation. 
We evaluated the instrumental  offsets using numerical 
routines designed for STIS by A. Marconi \citep{marconi02}, finding that
the offsets  have absolute values in the range 
4 -- 9~~km\,s$^{-1}$, being generally 
larger for zones closer to the star and for 
the slit positions S2 and S6, for which the illumination gradient is steeper. 
Once the spurious contribution has been determined, 
we have  corrected the instrumental deformation 
by applying opposite offsets 
to the affected pixel rows in each lateral spectrum.

The velocity shifts across the jet were then 
determined by applying both multiple Gaussian fits 
to the (sometimes complex) line profiles, and 
cross-correlation routines that analyse the overall 
displacement of lines.
The uncertainty in the pipeline wavelength calibration 
across exposures due to effects other than uneven illumination is 
reported to be $\pm 5$\,km\,s$^{-1}$ 
(HST Data Handbook for Cycle 10). 
This value is similar to the effective accuracy of the Gaussian fit.
The cross-correlation procedure, which measures
directly the displacement of the profiles, is also accurate to 
$\pm 5$\,km\,s$^{-1}$ for velocity {\em differences}.

\section{Results: possible evidence for rotation of the flow} \label{rot}
  
Here we concentrate on the innermost portion of the flow, i.e. the first 
0\farcs 5, which is inclined at 
$\approx$ 38\degr\, with respect to the line of sight \citep{eismundt98}.
In Fig.~\ref{vpeaks} we present the radial velocity 
values of the LVI/MVI emission
peak across the jet, estimated with  Gaussian fits and 
corrected for the STIS offsets, in each of the four regions 
described above. The velocity scale is negative, since 
the flow is blueshifted. Along the abscissa we have put the 
relative slit position in steps of 0\farcs 07
(corresponding to about 10\,AU  for Taurus).
In this plot, the star is on top, and the axis of the flow 
is in the middle of the figure. 
Apart from the systematic increase of radial velocity 
towards the central axis, there is a clear tendency for 
the northwestern side of the flow (positive positions, 
corresponding to slits S5, S6 and S7) to appear less blueshifted
than the southeastern side.  This result is better 
illustrated in Fig.~\ref{vshifts} (open symbols), where we 
plot the {\em difference} between the radial velocity values displayed in 
Fig.~\ref{vpeaks}, for slit positions located symmetrically 
with respect to the central one.
Assuming in turn that the emission from the flow
is axially symmetric, the value obtained for any  
position pairs (S1 -- S7), (S2 -- S6), (S3 -- S5) represents
the global velocity shifts between the two sides
of the flow.   
The above findings are confirmed by the velocity shifts determined by 
cross-correlating the line profiles for the same position pairs. 
The results of this 
procedure appear in  Fig.~\ref{vshifts} as filled symbols, 
and should be considered
as more accurate, since they are independent of the
shape of the line profile. The agreement is quite
good, with major discrepancies arising solely 
for the [OI]$\lambda$6300 line, as expected since this line is only
partially visible on the CCD \citep{bmresc00}.

In almost all positions, and for 
all lines, the velocity difference is {\em negative}, 
with values ranging between  5 to 20\,km\,s$^{-1}$.
In other words the observations 
are consistent with the southeastern side of the flow 
approaching us, and the northwestern side moving away.
The only domain where the general trend is not followed is 
for (S1-S7) in region I. 
This discrepancy is probably due to the faintness of the lines
in this region at large axial distances (see Sect.\,\ref{disc.2}). 
We also note that 
major positive shifts are detected for 
[SII]$\lambda$6716 in regions I and II but again this line is faint 
as the density is high \citep{brmes02} in these regions. Furthermore 
the [SII]$\lambda$6716 line profile is contaminated by its proximity 
to the LiI\,$\lambda$6707 line.

\begin{figure}
\epsscale{0.9}
\plotone{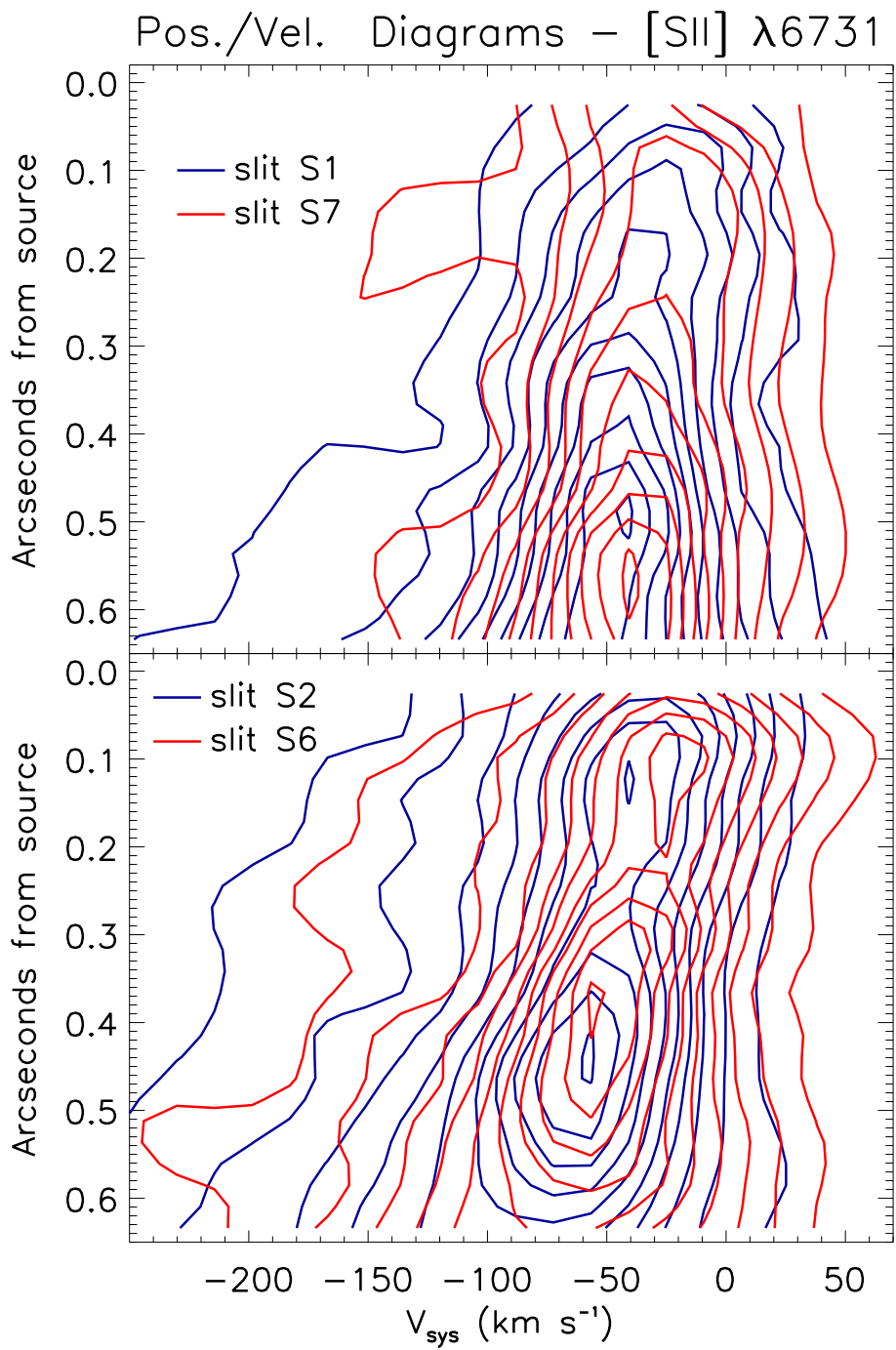}
\caption{[SII]$\lambda$6731 position-velocity diagrams for 
the region from 0\farcs 02 to 0\farcs 6 from the source. In the top 
panel contours for equally axially displaced slits S1 (blue) and 
S7 (red) are compared. The bottom panel shows S2 (blue) and S6 (red). 
The contour spacing are 10\% of the highest contour levels, 
that correspond to $2.7~10^{-14}$,  $2.3~10^{-14}$,
$4.6~10^{-14}$ and  $3.4~10^{-14}$ erg s$^{-1}$ cm$^{-2}$ \AA$^{-1}$ 
arcsec$^{-2}$ for S1, S7, S2 and S6 respectively. All velocities
are with respect to the systemic velocity of DG~Tau, and correction for uneven
slit illumination has been applied.  
\label{s6731pv}}
\end{figure}

\begin{figure}
\epsscale{0.9}
\plotone{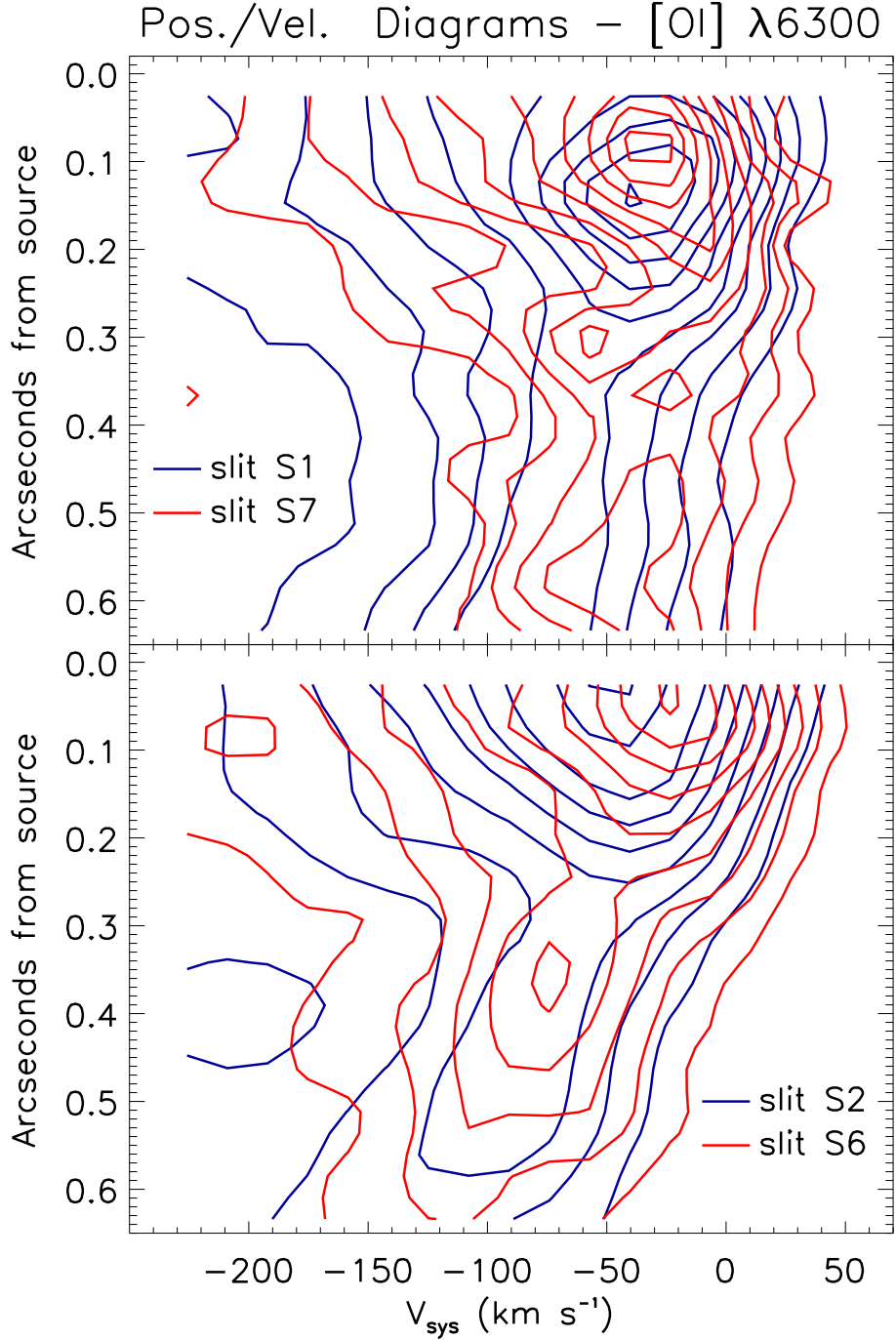}
\caption{As in Fig.\,\ref{s6731pv} but for the [OI]$\lambda$6300 line.
Highest contour levels correspond to  $3.3~10^{-14}$, $3.3~10^{-14}$,
$9.0~10^{-14}$ and $6.6~10^{-14}$ erg s$^{-1}$ cm$^{-2}$ \AA$^{-1}$ 
arcsec$^{-2}$ for S1, S7, S2 and S6 respectively.
\label{o6300pv}}
\end{figure}

\begin{figure}
\epsscale{0.9}
\plotone{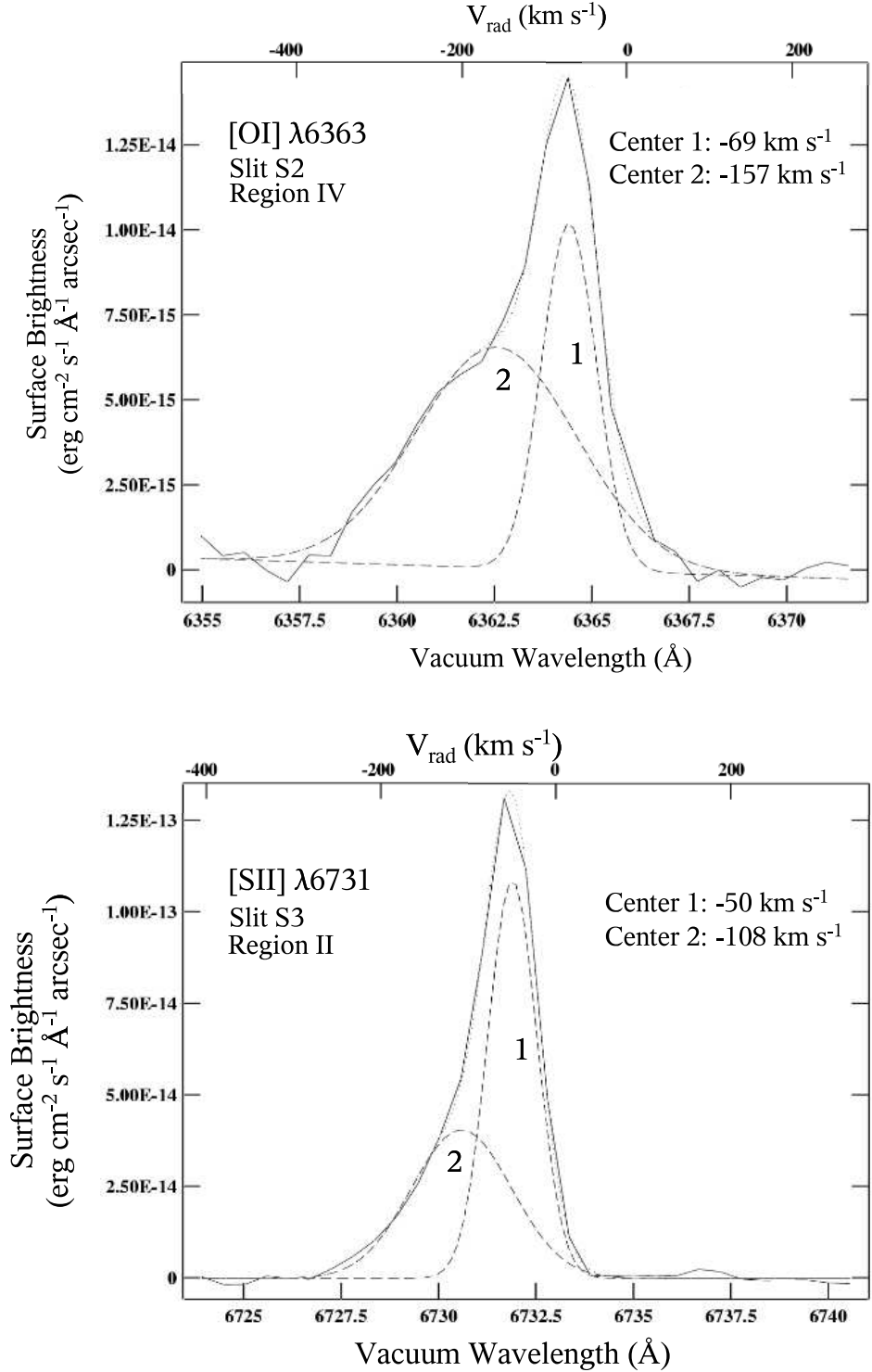}
\caption{Example of multiple Gaussian fits  
for the [OI]$\lambda$6363 line in the spectrum from slit S2 and in zone IV
(top panel) and for the  [SII]$\lambda$6731 line, in zone II and spectrum S3.
The analysis presented in the paper is based on 
the values obtained for the component
at low/moderate velocity, labelled with ``1'' in this Figure
\label{profiles}}
\end{figure}

Further illustrations of the lateral global velocity shifts across 
the flow are shown in Figs.\, \ref{s6731pv} and \ref{o6300pv}. 
Fig.\, \ref{s6731pv} contains superimposed position-velocity diagrams for the 
[SII]$\lambda$6731 line in the region between 0\farcs 02 to 0\farcs 6 from the 
source. In the top panel contours for the slit pair S1 and S7 are compared 
while the bottom one contrasts S2 and S6. All velocities
are with respect to the systemic velocity of DG~Tau, and the contour level
spacings are in intervals of 10\% of each peak value (indicated
in the figure captions).
The velocity shift between the southeastern and northwestern 
sides of the flow can be clearly seen. The 
corresponding diagrams for the [OI]$\lambda$6300  line are displayed 
in Fig.\, \ref{o6300pv}.
In Fig.\, \ref{profiles} we further show two typical line profiles, 
which are horizontal cuts of the position-velocity diagrams 
in selected positions. The Figure also illustrates the adopted 
multiple Gaussian fit. 
Our kinematical analysis refer to the well defined low-moderate 
velocity component labelled with ``1'' in the figure.

To check whether the observed differential Doppler shifts between the 
southeastern and northwestern parts of the flow contain any possible
systematic effect, we used the same procedures on an emission line
that should not produce any apparent differential shift.
H$\alpha$, at the stellar position, has three components:
an unresolved blueshifted wing at about -230\,km\,s$^{-1}$, 
that probably represents the base of the jet seen in projection, 
and two other components which are produced
at, or very close to, the photosphere. One is 
at approximately the systemic velocity
and the other is a red wing, at about +75\,km\,s$^{-1}$, attributed to
magnetospheric accretion (\citet{edwards97}).
As this emission is not resolved spatially in our spectra, 
it should not show any differential shift even if its 
formation region is rotating.
The result is illustrated in Fig.~\ref{hashift}: 
our analysis shows mean differential velocity shifts of almost zero.
This result strongly 
endorses our claim that the observed differential Doppler shifts in 
the other lines are not a result of systematic instrumental or
illumination effects.

\begin{figure}
\epsscale{0.9}
\plotone{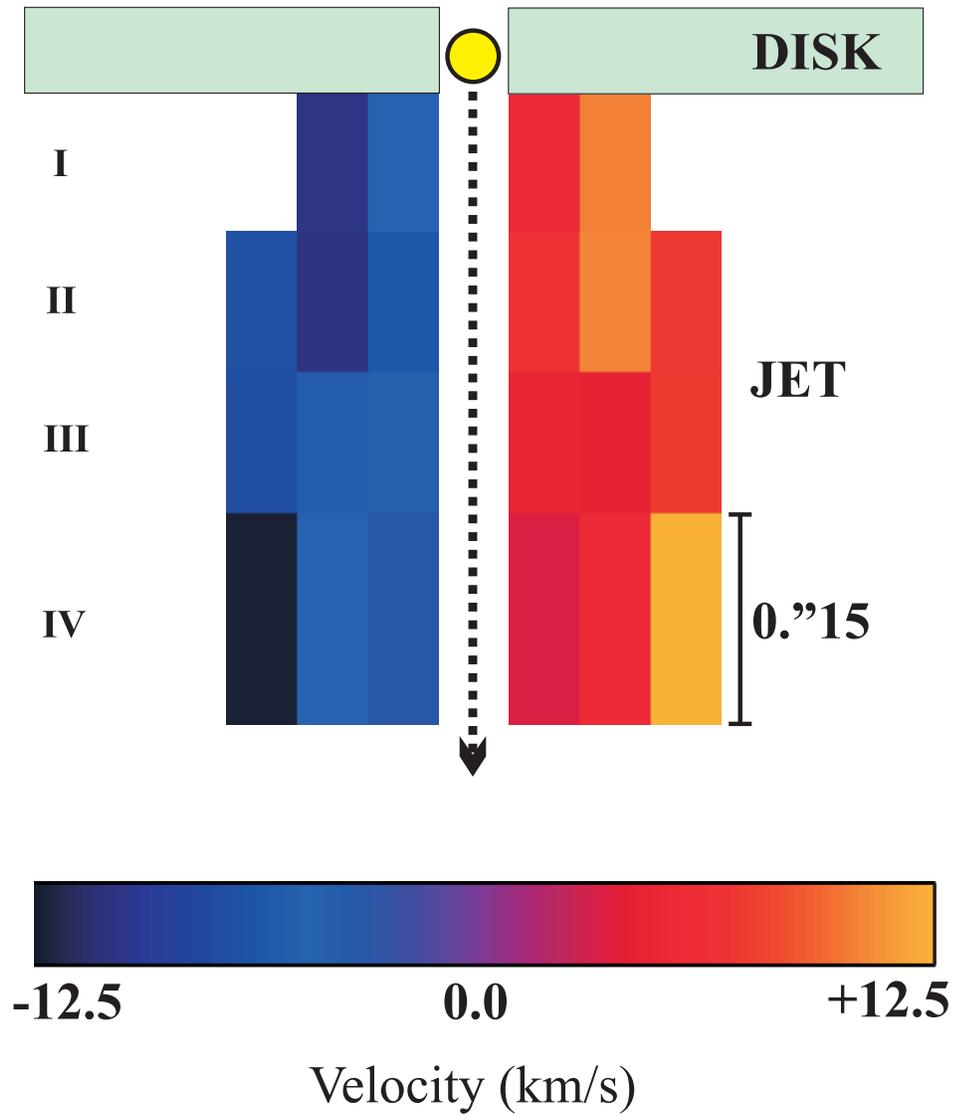}
\caption{Schematic map of the radial velocity shifts in 
the southeastern and nortwestern sides of the flow (projected 
onto the plane of the sky), derived from 
Table 1. Here the
flow is assumed to  be axially symmetric.  
The velocity scale is linear.
\label{spinmap}}
\end{figure}

If the dectected radial velocity differences 
are interpreted as rotation, 
they would imply that the flow is rotating clockwise looking 
from the tip of the flow toward the source. 
In Table\,\ref{tbl-1} we list 
the observed radial velocity shifts, averaged over the values 
found using the cross-correlation routines (i.e.\ filled symbols 
in Fig.~\ref{vshifts}) for the various lines.
In Figure~\ref{spinmap},
we represent the radial velocity shifts schematically, assuming an 
axially symmetric flow. \notetoeditor{This figure is in colours. 
Its quality may result not optimal on the screen, but it's OK 
when printed on paper.}  Here 0\farcs 1 corresponds to about 23 AU when 
{\em de-projected} onto the jet meridian plane. 
Blue and red colors indicate negative and positive projected  
velocity differences w.r.t.  the systemic velocity of the star. 
The velocity scale is linear.
Interpreting the velocity shifts as rotation, and
given the inclination angle of the flow, the toroidal velocity of the emitting 
regions would be between 6 and 15\,km\,s$^{\rm -1}$, depending on position.
For comparison, the Keplerian velocities in the disk, calculated 
for M$_{\star} = 0.67$ M$_{\odot}$ \citep{heg95} are 
$|7.8|$, $|5.5|$, and $|4.5|$\,km\,s$^{\rm -1}$ at 
9.8, 19.6 and 29.4 AU from the star, respectively.  
The latter velocities are given as 
absolute values, since the sense of rotation of the disk 
is not yet determined for this star. 
See further discussion in Sect.\,\ref{disc}.

Finally we should emphasise that there are some detailed radial 
velocity features which cannot be explained by rotation alone. For
example, close to the jet, the emission in S7 is faster than S6 but 
farther out the reverse is true (see Fig.~\ref{vpeaks}). Such 
discrepancies might well be attributed to clumpiness in the jet, 
that would be propably revealed with  higher angular resolution.
In any event it is likely that the
jet emission is  made up of several overlapping shocks.
Also, alternative interpretations to the observed systematic
velocity asymmetries could involve non-uniformities
in the jet environment, that may cause the bow-shock wings to be angled 
differently, and/or the jet-ambient entrainment to have different properties
on the two sides of the jets. Clearly, studies of more flows
are required. 

\section{Discussion} \label{disc}

\subsection{Further possible instrumental effects}\label{disc.1}

\begin{figure}
\epsscale{0.9}
\plotone{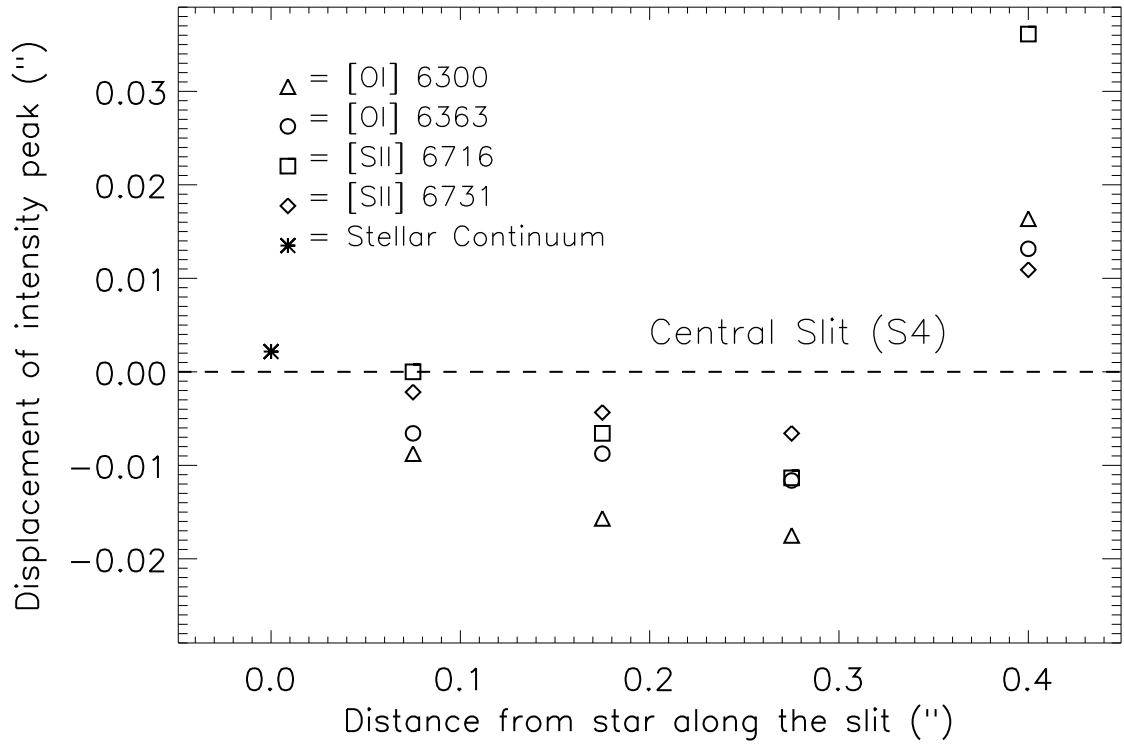}
\caption{Estimate of the spatial displacement 
of the intensity peak (for the LVI) with respect to the position 
of the central slit S4 (dotted line). See text for details.
\label{angles}}
\end{figure}

To test our interpretation of the transverse radial velocity differences 
in terms of rotation of the flow,
we have to  exclude other possible causes that could produce such an effect.
One source of error might be the 
orbital motion of the HST, since the linear speed of the spacecraft with
respect to the Earth is about
7~km s$^{\rm -1}$. Such an effect can however be ruled out, since 
the exposure time of each spectrum is distributed  
over the orbital period of the spacecraft. 
Another way apparent transverse radial velocity differences can
be produced is if the sequence of slit positions is not centered on the 
jet axis.  
For example let us suppose that the sequence is slightly offset
to the northwest. Even if the blueshifted flow from DG~Tau is 
{\em not} rotating and perfectly 
axially symmetric, apparent velocity differences, with the same sign 
as we observe, would be inferred if the velocity of the flow decreases 
with increasing distance from the central axis.
Such an effect, however, would only be relevant if the misalignment
of the slit is about 0.1 arcsec.
To determine whether any such offset was present (an unlikely scenario as we 
had used the STIS ACQ/PEAK option to peak up on the source), we determined 
the precise position and orientation of the slits by analyzing, with 
Gaussian fits, transverse intensity profiles in reconstructed images 
of the initial jet channel.
In particular, we used images in the [SII] and [OI] lines \citep{bmresc00} 
as well as images of the stellar continuum.  
Assuming the brightness distribution is axially
symmetric, the displacement of the emission peaks with respect to 
the central slit should indicate the relative position of the latter (and
hence, of the system of parallel slits) with respect to the flow axis.
The results are displayed in Fig.\,\ref{angles}.
It turns out that the central slit 
is (not surprisingly) well positioned on the star, 
within a small fraction of the slit width,
but that the slit orientation is slightly rotated with respect to the 
flow axis (as defined by the intensity distribution) by about 3\degr. Note,
however, that in the last position (i.e. Region IV) the peak intensity
is located on the opposite side of the slit, probably an effect of the 
opening of the $`$bubble' \citep{bmresc00}. In any event, such a slight 
rotation will produce only a marginal asymmetry in velocity
(estimated to be below 20\% of the detected values).  
Moreover, if this slight tilt was important, we would expect
{\em larger} velocity shifts toward the axis of the jet, contrary
to what we observe. Thus the small tilt of the slit can 
neither reproduce the   
magnitude nor the type of velocity offsets that we detect.

If we are indeed observing rotation of the flow, 
then it should be in the same sense as  in the
circumstellar material of the accretion disk and/or the parent 
molecular cloud. Unfortunately, we do not have any clear indications 
which way DG~Tau's disk is rotating
(A. Sargent, 2001, priv. comm.), while the 
large-scale molecular environment of the star 
appears to be characterized by a combination
of infall and outflow rather than rotation
\citep{kitamura96}. 

\subsection{Comparison with theoretical predictions} \label{disc.2}

A full comparison of the observed velocity values 
with existing models would require 
the construction of a simulated spectrum with excitation and kinematic 
parameters provided by the model. Unfortunately the peripheral regions of the 
outflow, i.e.\ the regions we are beginning to resolve, are generally less
accurately modeled than the axial portions.
That said, we can try to compare our results with general
theoretical predictions, assuming that the velocity shifts we observe 
are indeed due to the presence of rotation.
Starting from the seminal work of \citet{bp82}, 
the popular magneto-centrifugal launching models 
have been developed by several groups,
the most widely known being the
the $`$disk wind' model (see the review by \citet{konigl00}) 
and the  $`$X-wind' model, described by \citet{shu95,shu00}.
Similar approaches are also  adopted by  \citet{cam97},
\citet{ferr95} and  \citet{LHF}.
What renders these models particularly attractive is 
the fact that the outflow extracts  
angular momentum from the protostar plus
disk system, thereby allowing the central object to accrete 
matter  up to its final mass. 
Without going into the details, we list here a few general 
properties that are relevant to our study.
\\ \indent
(i) The wind is launched along magnetic field lines 
(in a $`$bead-on-a-wire' fashion) from a region 
of the disk between a few stellar radii (X-wind) and a few AU 
(or more, disk wind).
For both cold and warm flows, a wind can be launched if 
at the footpoint $r_0$ the inclination of the magnetic/flow 
surface with respect to the plane of the disk is less than 60\degr .
\\ \indent
(ii) The surface at which the poloidal 
velocity, $v_p$,  equals the characteristic poloidal 
Alfv\`en speed 
is called the Alfv\`en surface (hereafter referred to as the AS). 
Here, the kinetic fluid energy 
becomes dominant, and the inertia of the fluid 
deformes the  shape of the field: a strong $B_{\phi}$ component 
is generated, which tends to collimate the jet (due to hoop stresses).
The distance from the axis at which the flow surface intersects the AS is
called the Alfv\'en radius, $r_A $. 
Since the fluid is accelerated
up to the AS, the distance $r_A - r_0$ represents the 
lever arm of the magnetic field.
Up to the AS, the flow slides along the lines of
force as if moving with solid body rotation. Hence below the AS
$v_{\phi} \propto r $, where $r$ is the radial distance from the rotation 
axis. Above the AS the flow moves along the magnetic surfaces conserving
its angular momentum, and hence here $v_{\phi} \propto r^{-1}$.
\\ \indent
(iii) For a $`$cold' flow (i.e.\ enthalpy negligible with respect to kinetic 
and 
magnetic energy), the terminal poloidal velocity along each magnetic/flow 
surface is given by
$
v_{p,\infty} \sim 2^{1/2} (r_A/r_0)v_K 
$,
where $v_K = (G M_{\star})^{1/2} r^{-1/2}_0 $ is the Keplerian 
velocity in the disk at the footpoint.
\\ \indent
(iv) 
If viscous stresses are negligible with respect to the magnetic torque,
the concept of global angular momentum conservation 
finds its formal translation in the simple relationship:
$
\dot{M}_{\rm jet}/\dot{M}_{\rm acc} \sim  
(r_0/r_A)^2,
$ 
where $\dot{M}_{\rm jet}$ and  $\dot{M}_{\rm acc}$ are the rates
of mass flow within the jet  
and accreted through the disk, respectively. 
\\ \indent
(v) The ratio  $r_A/r_0$ varies between the models, but 
it is generally found to be in the range 2 -- 5, 
both on analytical and numerical grounds. As a consequence,
$\dot{M}_{\rm jet}/\dot{M}_{\rm acc} \sim 0.1$, in agreement
with observed values for various jets
(including the DG Tau outflow itself, see \citet{brmes02}).

If we now turn to our results, we can first check if the values obtained  
for the toroidal velocities are in the right range for 
magneto-centrifugal acceleration. 
In \cite{brmes02} we estimate that 
$\dot{M}_{\rm jet}/\dot{M}_{\rm acc} \sim 0.1$ for this flow.
Thus from (iv) and (v) we deduce that on average 
$r_A = 3\,r_0$. We will assume for simplicity 
that the same ratio holds for any
magnetic/flow surface rooted in the disk. 
From Fig.~\ref{vpeaks} we deduce that the average 
radial velocity measured on the jet symmetry axis for the 
resolved flow is -70\,km\,s$^{-1}$,
which de-projected along the flow surfaces gives $v_p =$ -80\,km\,s$^{-1}$
(see \citet{brmes02} for details). If we take this value as the 
terminal velocity $v_{p,\infty}$ of the velocity interval we are looking at,
and we assume a cold flow, we deduce from (iii) that 
$v_K \sim$ 18\,km\,s$^{-1}$.
Hence the launching region for this zone is centered 
at $r_0 \sim$ 1.8\,AU. Let us now follow the fluid particle 
(the $`$bead') in its movement from the footpoint $r_0$ along 
the representative magnetic line (the $`$wire') rooted at $r_0$. 
According to (ii), matter acquires its maximum linear rotational 
velocity at the AS, which in the
present example crosses the relevant flow surface at $r_A \sim 5.4$\,AU
from the axis (and hence below the resolution limit of HST itself). 
Here $v_{\phi,max} \sim 54$\,km\,s$^{-1}$. 
After the passage through the AS, the 
flow starts to lag behind, conserving its angular momentum.
The expected value of $v_\phi$ at a distance $r$ from the axis 
depends on model details, and 
in particular upon the shape of the magnetic surfaces. 
We can estimate, however, that angular momentum conservation above 
the AS would yield $v_\phi =$ 29, 15 and 10\,km\,s$^{-1}$ 
if the magnetic surface is responsible for the flow
at 10, 20 and 30 AU from the axis, respectively.
Thus we see that the toroidal velocities we have inferred in 
Section~\ref{rot} from our observations  for 
similar distances from the rotation axis (and well above the 
expected location of the AS) are in very good agreement 
with the range provided by  simple and very general theoretical arguments.

We can take a step further by estimating the angular momentum flux
$\dot{J}_{w,lm}$ in the low to moderate velocity region of the flow. 
This is given by
\[
\dot{J}_{w,lm} = \int^{r_{\rm out}}_{r_{\rm in}} \rho~ v_p~ r v_{\phi}~
2 \pi r ~dr,  
\]
where $\rho$ is the total mass density and $r_{\rm in}$ 
and $r_{\rm out}$ are the inner and outer radii, at a given height
above the disk, of the partially hollowed-out cone containing the low to 
moderate velocity flow. Since the detailed variation of these  
quantities across the jet are not available from 
observations, we will make a simplified estimate 
using mass loss rates and flow radii (for the aforementioned velocity 
intervals) from \citet{brmes02}. The regime 
examined, isolated on the basis
of the shape of the line profiles, is the combination of the LVI 
($v_{\rm rad} =$ +55 to -70\,km\,s$^{-1}$) and
MVI ($v_{\rm rad} =$ -70 to -195\,km\,s$^{-1}$).
For these intervals, the mass loss rates are 
calculated to be 8.3\,10$^{-8}$\,M$_\odot$\,yr$^{-1}$ (LVI), 
and 5.1\,10$^{-8}$\,M$_\odot$\,yr$^{-1}$ (MVI).
Thus, assuming that in both the LVI and the
MVI the average toroidal velocity is $\bar{v}_\phi = 15$\,km\,s$^{-1}$ 
we can estimate the angular momentum flux as 
\[
\dot{J}_{w,lm} = \bar{v}_\phi (\dot{M}_{\rm LVI}\bar{r}_{\rm LVI} +
\dot{M}_{\rm MVI}\bar{r}_{\rm MVI})~\sim~3.8\,10^{-5}
\,{\rm M_\odot\,yr^{-1}\,AU\,km\,s^{-1}},
\]
where $\bar{r}_{\rm LVI}=$ 21.6\,AU and $\bar{r}_{\rm MVI}=$ 14\,AU
are the average radii of the zones.

For comparison, we calculate the 
angular momentum loss rate $\dot{J}_{D,r_0}$ that the disk must 
sustain at the footpoint $r_0$ in order to let the matter move inward 
with the observed $\dot{M}_{\rm acc} \approx$ 2\,10$^{-6}$ 
M$_\odot$~yr$^{-1}$ \citep{heg95,brmes02}. 
It turns out that $\dot{J}_{D,r_0} \sim $ 6.5\,10$^{-5}$ 
M$_\odot$\,yr$^{-1}$\,AU\,km\,s$^{-1}$. 
Thus the flux carried away by the wind in the 
low to moderate velocity intervals amounts to
about 60\% of the necessary angular 
momentum loss for the disk at its footpoint.
As we used several simplifying assumptions, the true angular momentum flux 
carried by the wind could 
be as much as 100\%\ of that transported by the disk. 
Note however that viscous stresses could also transfer angular 
momentum outwards through the disk itself. 
If such stresses are significant, the simple relationship in (iv) 
becomes invalid.  
Also, we note that the above considerations
assume steady state conditions. On the other hand, 
DG Tau is known to be highly variable, and variability obviously
introduces further uncertainties.

\section{Conclusions} \label{conc}

Many features of the collimated HH jets associated with star formation
are still unexplained, especially the origin of the jets themselves. 
Models for the launching of flows still lack 
observational constraints, due to the small size of the acceleration 
zone and the fact that the central sources are often heavily embedded. 
In order to shed some light on this area, we have taken and 
analysed a set of seven high angular resolution  ($\leq 0\farcs1$)  
HST/STIS spectra of the outflow from the T Tauri star DG~Tau.   
Here we report possible evidence for rotation in this dataset: 
rotation, in fact, is a fundamental ingredient
for the modelling of the acceleration of outflows and of the interplay
between accretion and ejection of matter in the framework of
the formation of a new star. 

From a detailed analysis of the line profiles in each spectrum,
and in four distinct regions of the initial part of the jet 
(within about 100\,AU from the star), 
we have found systematic differences in the radial velocity
of the lines for each pair of slits 
displaced symmetrically with respect to the 
jet axis. The values, obtained with multiple Gaussian fitting and/or
cross-correlation routines, have been  corrected for the wavelength offset
produced by the uneven illumination of the STIS slit. 
Other possible instrumental effects that may contaminate
the data are either absent or unimportant. 
According to our results, and under the assumption that the flow is 
axially symmetric, the southeastern side of the blueshifted jet appears to 
move 
toward the observer faster than the 
corresponding northwestern side, and the average value found for the 
difference 
is about 10\,km\,s$^{-1}$. A detailed map of the shifts is available in 
Fig.\,\ref{spinmap}, derived from the values in Table\,\ref{tbl-1}.  
If we interpret these findings in terms of rotation
of the flow, they would imply that the jet is rotating clockwise 
looking from the flow tip towards the source. 
Taking into account the inclination of this system 
with respect to the line of sight, 
one would derive apparent toroidal velocities of around 6 to 15\,km\,s$^{-1}$
at a few tens of AU from the axis, and between 20 to 90\,AU above the disk 
plane. These velocities would be in the range predicted by MHD theories.
We also estimate  the angular momentum flux in the low to medium velocity 
regime to be about 3.8\,10$^{-5}$\,M$_\odot$\,yr$^{-1}$\,AU\,km\,s$^{-1}$.
This could amount to 60\% of the angular momentum that the disk has
to loose per unit time at the footpoint in order to accrete at the observed 
rate. Our findings may constitute 
the first detection of rotation in the initial channel
of a jet flow, although at the present stage we cannot exclude
that non-uniformities in the jet and/or in its environment  
cause  the observed asymmetries. More observations of this and other
outflows are required. If the future 
studies will confirm the presence of rotational motions
at the base of the flows,
it would represent an important validation of magneto-centrifugal
models for the launching of both galactic and extra-galactic jets.





\acknowledgments
We are indebted to A. Raga, S. Shore and A. Marconi 
for their comments and their help in the analysis
of the  line profiles. Thanks are also due to an anonymous 
referee for their critical appraisal of the original manuscript 
and helpful suggestions.  Finally FB wishes to acknowledge ESA 
for financial support and 
T. Lery and Z.-Y. Li for fruitful discussions about MHD models.



\appendix
\section{Correction for uneven slit illumination}\label{app1}

\begin{figure}
\epsscale{0.9}
\plotone{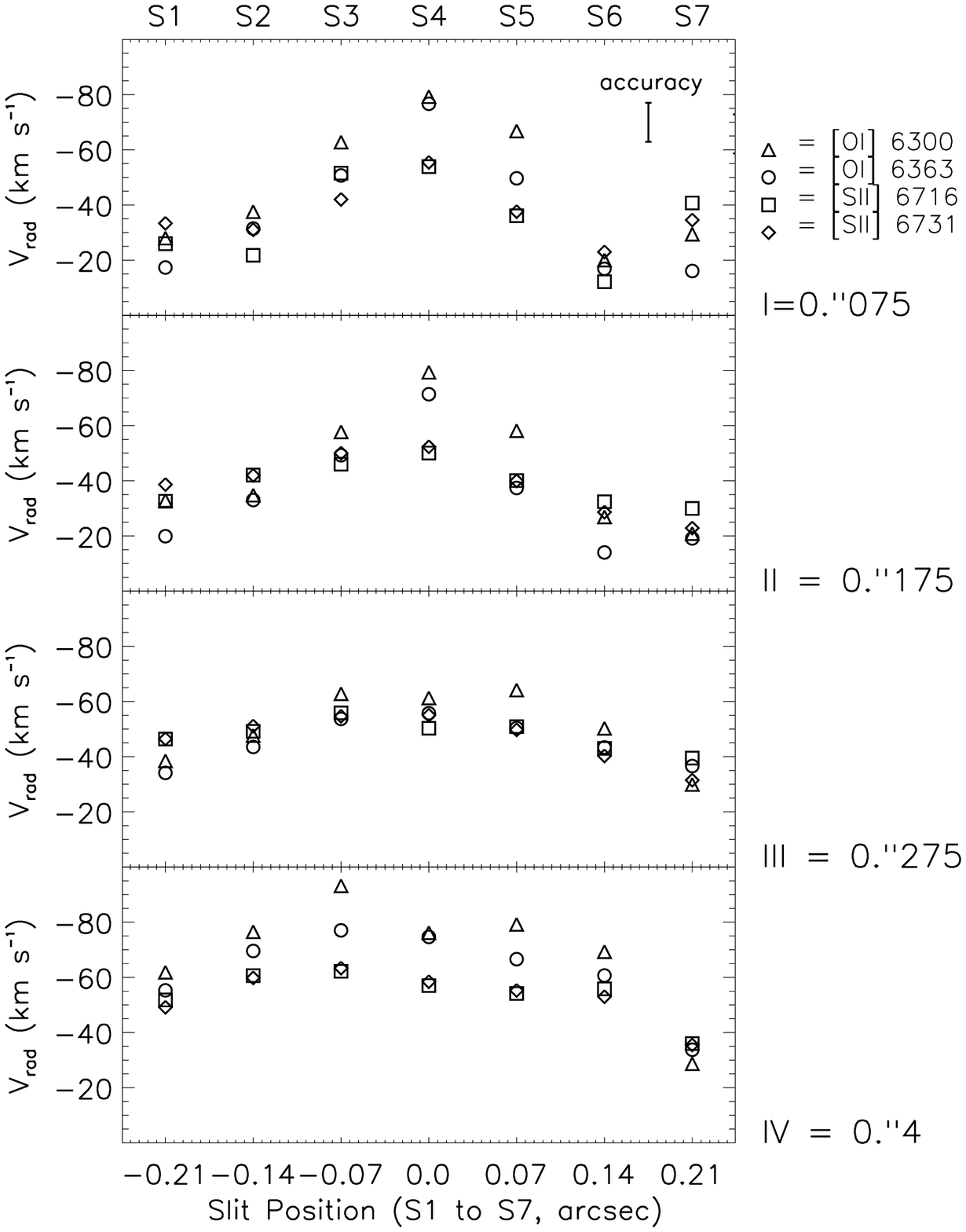}
\caption{Radial velocity values 
of the line peak at low/moderate velocity,  
for all slits across the jet
(S1 to S7, from the southeastern to the northwestern border of the jet
respectively), and in four distinct regions along the flow
(I, II, III and IV, see figure for distances from the source).
The values are derived from multiple Gaussian fitting of the line profiles,
and are corrected for the effects of uneven slit illumination.
The error bar illustrates the typical accuracy of the Gaussian fit
combined with the one of the STIS pipeline calibration. 
\label{vpeaks}}
\end{figure}

\begin{figure}
\epsscale{0.9}
\plotone{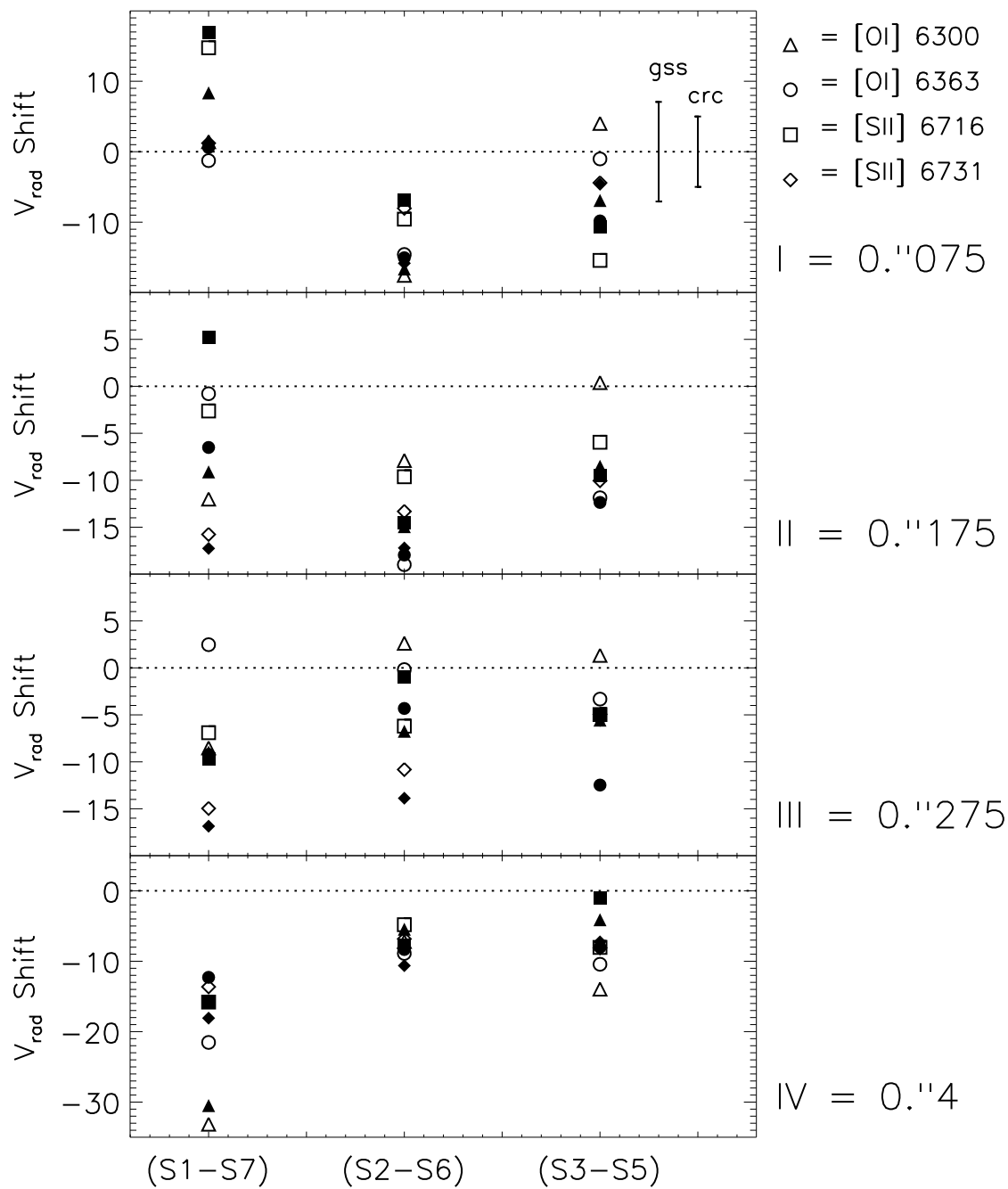}
\caption{Radial velocity differences between
slit positions displaced symmetrically 
with respect to the central slit. Units are km\,s$^{-1}$.
Open symbols: difference of the values displayed in Fig.\,\ref{vpeaks}
(error bar labeled with `gss').
Filled symbols: higher accuracy values obtained by cross-correlating 
the associated line profiles (error bar label: `crc').
In almost all positions the velocity difference is negative,
indicating a net relative motion of the southeastern 
side of the flow with respect to its northwestern side.
\label{vshifts}}
\end{figure}

\begin{figure}
\epsscale{0.9}
\plotone{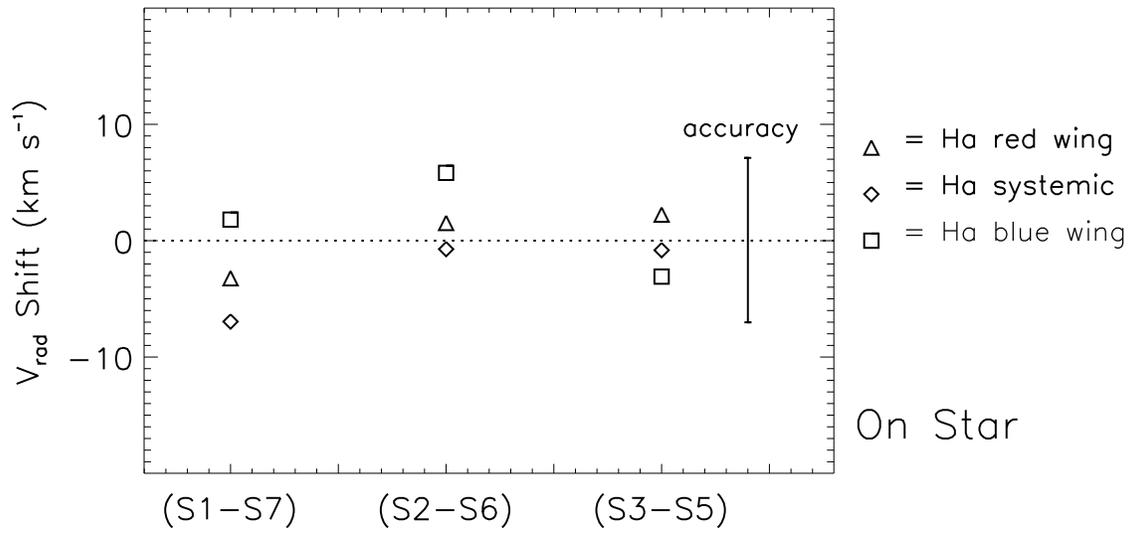}
\caption{Velocity shifts, as in Fig.\,\ref{vshifts}, but for 
the H$\alpha$ line at the stellar position. The systemic 
(diamonds), redshifted (triangles) and blueshifted 
(squares) components of this line have been 
analysed. All the components are unresolved, 
and are not expected to show any shift when 
corrected for the STIS offset. Note that the observed shift is very limited 
illustrating the high sensitivity of our technique.
\label{hashift}}
\end{figure}

Following \citet{marconi02} 
the light distribution at a point $(x,y)$ on the focal 
plane of the telescope can be expressed as 
$ \Phi(x,y,v)=\int \int^{+\infty}_{-\infty} dx^{\prime} dy^{\prime}
P(x-x^{\prime},y-y^{\prime})
I(x^{\prime},y^{\prime})
\phi(v-u(x^{\prime},y^{\prime})) $, where 
$I$ is the total intensity at a point 
$(x^{\prime},y^{\prime})$ in a reference frame $X',Y'$ on the sky,
$\phi$ is the intrinsic velocity profile 
centered at the radial velocity $u(x^{\prime},y^{\prime})$, 
and $P$ is the instrumental PSF. On the focal plane, the slit is
aligned with $Y=Y^{\prime}$, its center is at $(x=x_0,y=0)$, 
and its width is $\Delta x$.
Past the slit and the spectrograph, the light is recorded 
on the detector, which combines  $x$ and  
$v$  into the coordinate $w=v+k(x-x_0)$, interpreted as
the measured velocity. In the latter, 
$k(x-x_0)$ is the spurious velocity contribution 
introduced from light entering off the slit center.
The coefficient $k$ is given by   
$k= \Delta w / \Delta \xi$, where $\Delta \xi = 0."05477$ 
is the STIS plate scale in the dispersion direction  \citep{STIS98}
in the considered wavelength range. The expected line profile 
$\tilde{\Psi}_i(w)$ at the pixel row $i$ of the detector 
is then obtained by integrating 
the light contributions across the slit and over the width $2\Delta y$
of the $i^{{\rm th}}$ pixel, and by convolving the result  
with the shape of the pixel in the dispersion direction,
a top-hat of width $2\Delta w$. Finally, the average velocity $<v_i>$
measured at pixel $i$ along the slit  can be expressed through
the first order momentum of $\tilde{\Psi}_i(w)$ with respect to $w$.
More precisely,   \citet{marconi02} show   that  
\begin{equation}
<v_i> = {2\Delta w 
\int^{x_0+\Delta x}_{x_0-\Delta x} dx
\int^{y_i+\Delta y}_{y_i-\Delta y} dy
\int \int^{+\infty}_{-\infty} dx^{\prime} dy^{\prime}
I(x^{\prime},y^{\prime})
P(x-x^{\prime},y-y^{\prime}) 
(u+k(x-x_0))
\over
\int^{+\infty}_{-\infty} \tilde{\Psi}_i(w) dw}.
\label{eq_offset}
\end{equation} 
It can thus be argued that if $I$ is not constant across the slit,
the term $k(x-x_0)$ will be weighted differently with varying $x$, 
and this can cause a  spurious velocity offset 
to appear at the detector. For a point source, and for the adopted STIS 
configuration, the spurious shift can be up to
about 25\,km\,s$^{-1}$. The effect is reduced,
but not eliminated, if the observed source is extended, as in our case. 
We estimated the wavelength offsets 
using numerical routines recently 
developed by A. Marconi \citep{marconi02}. 
For a given slit width and slit position with respect to the source, 
the routines calculate the average velocity measured at the detector through 
Eq.~\ref{eq_offset}, starting from a model intensity and velocity field 
$I(x^{\prime},y^{\prime}),~u(x^{\prime},y^{\prime})$.
The adopted  PSF is the one generated by TinyTim
at 6700 \AA, and convolution is done using a {\it Fast Fourier Transform}
algorithm. To estimate the spurious offset in each line/position, 
we used a model intensity distribution derived by combining the surface 
brightness in the LVI and MVI intervals  
shown  in \citet{bmresc00}. For the purpose of the calculation, we have then 
assumed a constant (arbitrary) velocity field 
$u(x^{\prime},y^{\prime})=\bar{u}$. 
The {\em difference} $(<v_j> - \bar{u})$ is thus the value of the instrumental offset 
to be determined.
It should be noted that in the real case in which 
the velocity field is not constant,  
the passage through the telescope and spectrograph
introduces an additional  deformation to the transverse profile of 
the velocity. In our case, however, this effect should not lead to
spurious velocity asymmetries  
with respect to the central axis, because the radial velocity, as well as
the intensity, is increasing (in absolute value) toward the axis on 
both sides of the jet (see Fig.~\ref{vpeaks}).

\clearpage

\clearpage 

\clearpage

\clearpage

\clearpage

\clearpage

\clearpage

\clearpage

\begin{deluxetable}{lccccc}
\tablecaption{Observed radial velocity shifts \label{tbl-1}}
\tablewidth{0pt}
\tablehead{
\colhead{Region} & 
\colhead{$d_s$ \tablenotemark{a}}   & 
\colhead{$d_j$ \tablenotemark{b}}   & 
\colhead{$v_{\rm rad}$ shift for (S1--S7)}&
\colhead{$v_{\rm rad}$ shift for (S2--S6)}&
\colhead{$v_{\rm rad}$ shift for (S3--S5)} \\
\colhead{ } &
\colhead{(\arcsec)} &
\colhead{(AU)} &
\colhead{(km\,s$^{-1}$)} &
\colhead{(km\,s$^{-1}$)} &
\colhead{(km\,s$^{-1}$) }
}
\startdata
I & 0.075 &17& +3.4\tablenotemark{c,d} & -15.8\tablenotemark{c} & -8.0 \\
II & 0.175 & 40 &-11.0\tablenotemark{c}  & -16.2   &  -9.9 \\
III & 0.275 & 62 &  -11.2 & -6.5 & -7.2 \\
IV & 0.4 & 91 & -19.1 & -8.0 & -5.1
\enddata
%
\tablenotetext{a}{Angular distance from the position of the star.} 
\tablenotetext{b}{Corresponding distance de-projected 
onto the jet meridian plane.}
\tablenotetext{c}{Value calculated excluding 
the velocity shift derived from the [SII]$\lambda$6716 line from the average
(see text).}
\tablenotetext{d}{In this position the forbidden lines are 
very faint and the positive value quoted has a large 
uncertainty (see text).}
\end{deluxetable}

\end{document}